\documentclass[preprint,showpacs,preprintnumbers,amsmath,amssymb]{revtex4}
\usepackage{graphicx}% Include figure files
\usepackage{dcolumn}% Align table columns on decimal point
\usepackage{bm}% bold math
\begin{document}
\title{Phase diagram of bismuth in the extreme quantum limit}
\author{Huan Yang$^{1}$, Beno\^{\i}t Fauqu\'e$^{1,2}$,Liam Malone$^{3}$, Arlei B. Antunes$^{3}$, Zengwei Zhu$^{1,4}$, Ctirad Uher$^{5}$ and Kamran Behnia$^{1}$}
\affiliation{(1)LPEM (UPMC-CNRS), Ecole Sup\'erieure de Physique et de Chimie Industrielles, 75005 Paris, France\\
(2)H. H. Wills Physics Laboratory, University of Bristol, Bristol, BS8 1TL, United Kingdom\\
(3)Laboratoire National des Champs Magn\'{e}tiques Intenses (CNRS), 38042 Grenoble , France\\
(4) Department of Physics, Zhejiang University, Hangzhou, 310027 China\\
(5) Department of Physics, University of Michigan, Ann Arbor, Michigan 48109, USA}
\date {June 17, 2010}

\begin{abstract}
 Elemental bismuth provides a rare opportunity to explore the fate of a three-dimensional gas of highly mobile electrons confined to their lowest Landau level. Coulomb interaction, neglected in the band picture, is expected to become significant in this extreme quantum limit with poorly understood consequences. Here, we present a study of the angular-dependent Nernst effect in bismuth, which establishes the existence of ultraquantum field scales on top of its complex single-particle spectrum. Each time a Landau level crosses the Fermi level, the Nernst response sharply peaks. All such peaks are resolved by the experiment and their complex angular-dependence is in very good agreement with the theory. Beyond the quantum limit, we resolve additional Nernst peaks signaling a cascade of additional Landau sub-levels caused by electron interaction.
\end{abstract}

\maketitle

Seventy years ago, an intensive study of the angular dependence of the de Haas-van Alphen effect in elemental bismuth\cite{shoenberg} led to the first experimental determination of the Fermi Surface in any metal. The deduced structure consists of one hole ellipsoid aligned perpendicular to the plane in which lay obliquely three slightly tilted electron ellipsoids (Fig.1a). Later investigations, carried out during the following decades, led to a quantitative description of the size, the orientation, and the position of these four ellipsoids\cite{smith,bhargava,edelman,liu}. When the magnetic field is aligned along the highest symmetry axis, known as the trigonal axis, a field of 9 T allows one to attain the quantum limit putting carriers in their lowest Landau level.

The Nernst response in this configuration has been recently found to sharply peak each time a Landau level empties, generating giant quantum oscillations\cite{behnia2,behnia1}. This particular Nernst profile, absent in two-dimensional systems, has been recently observed in bulk graphite \cite{zhu}. It points to electronic degrees of freedom \emph{along} the magnetic field as a source of transverse in-plane thermoelectric response. In contrast to its two-dimensional counterpart\cite{jonson}, the theoretical description of the Nernst effect in a three-dimensional system in the vicinity of the quantum limit has only recently begun to be explored\cite{bergman}.

The extension of the Nernst measurements to fields as high as 33 T led to the observation of three Nernst peaks above 9 T\cite{behnia1}, where no more Landau level crossing were expected. The latter observation raised the issue of collective effects in a three-dimensional electron gas at high magnetic field where Coulomb interaction and associated many-body instabilities are expected to become significant\cite{halperin} and the adequacy of the band picture, which treats electrons as non-interacting entities, remains an open question. Several recent theoretical studies explored the limits on the survival of the Fractional Quantum Hall Effect in the presence of finite interlayer coupling\cite{burnell,levin,alicea2}.

Subsequent experimental work found that in strong magnetic field, both bismuth\cite{li,fauque1,fauque2} and Bi$_{1-x}$Sb$_{x}$\cite{banerjee} are far from featureless. Transport and thermodynamic coefficients were found to display unexpected anomalies beyond the quantum limit.  In particular, angular-dependent torque magnetometry measurements detected a field scale quite distinct from the Nernst anomalies and identified this field scale as a phase transition implying the massive Dirac electrons of the three electron ellipsoids\cite{li}. However, two independent set of theoretical calculations\cite{alicea,sharlai} found that the one-particle spectrum of bismuth for a field oriented close to the trigonal axis is not trivial. Charge neutrality in a compensated system implies strict equal density of holes and electrons as the magnetic field is swept. The intricate topology of the Fermi surface and the large Zeeman energy conspire to generate a complex phase diagram. According to these calculations\cite{alicea,sharlai}, the Landau level crossing of the three electron ellipsoids occur at a field which sharply shifts with angle as the field is slightly tilted off the trigonal axis. The field scale found by torque magnetometry\cite{li} follows the theoretical angular dependence of the field scale associated with the intersection of the Landau level of an electron pocket and the Fermi level. On the other hand, according to the same theoretical calculations, if the field is strictly oriented along the trigonal axis no other Landau level crossing is expected beyond 10 T. Thus, the torque anomalies were expected in the one-particle picture\cite{alicea,sharlai}, but not the three high-field anomalies observed in the Nernst response\cite{behnia1}. It was suggested\cite{sharlai} that they could be attributed to the Landau levels of the electron ellipsoids assuming a small misalignment of a few degrees. Due to the absence of \emph{in situ} orientation of the crystal in the Nernst experiment\cite{behnia1}, this possibility could not be ruled out.

In this paper, we present the first study of angular-resolved Nernst effect in bismuth and map the angular variation of the Nernst peaks across the quantum limit. Both sets of high-field anomalies found in the two previous studies\cite{behnia1,li} are present and easily distinguishable in our data.  According to our findings, the one-particle picture is relevant yet insufficient. The Nernst peaks caused by the emptying of a Landau level, closely follow the theoretically expected angular path. On the other hand, there are three unexpected Nernst peaks which cannot be associated with any Landau level.  According to the non-interacting picture, the Nernst peaks occur at a field when the Fermi energy is located \emph{between} Landau levels and no sharp feature in the density of states is expected. This result points to collective effects enhancing entropy per carrier at three particular magnetic fields between 9 T and 28 T.

\section{Results}

\subsection{Low field measurements}
Fig. 1 presents the result of an angular-dependent study performed at T=0.49 K and up to 12 T.  As previously reported\cite{behnia2}, the Nernst response shows quantum oscillations with a period (0.147 T$^{-1}$) matching the cross section of the hole ellipsoid. In this configuration, the Nernst response, like other transport properties such as resistivity\cite{bompadre} and the Hall effect\cite{fauque1} is dominated by holes. The lower mobility of carriers traveling perpendicular to the trigonal axis on an electron ellipsoid\cite{hartman} makes them less visible in a transport experiment. On the other hand, torque magnetometry\cite{li,fauque1}, a probe of anisotropic magnetization is dominated by the electron ellipsoids. The latter are much more anisotropic than the hole ellipsoid and their diamagnetic response is enhanced by their Dirac-like dispersion.

As the field was tilted in the (trigonal, binary) plane of the crystal, the main peaks barely moved. But the structure visible between large peaks rapidly evolves with tilt angle, $\theta_{1}$, as seen by arrows tracking one of the smaller Nernst peaks. Fig. 1c presents a color map of the Nernst response in the (B, $\theta_{1}$) plane together with the position of the smaller Nernst peaks. One can clearly distinguish between two different field scales. The first are quasi-horizontal field scales, caused by the passage of hole landau levels as previously identified\cite{behnia2,sharlai}.  The second group of field scales are less prominent in the data and present a very sharp angular variation. Their coordinates in the (B, $\theta_{1}$) plane coincides with the torque anomalies in the same field window\cite{li} and we thus deduce that they are by the passage of the Landau levels of the electron pockets in agreement with the predictions of both sets of theoretical calculations\cite{sharlai,alicea}. In particular, the electron spectrum calculated by Alicea and Balents \cite{alicea} appears to be in quantitative agreement with our data. Note the presence of a small Nernst peak (marked by a green arrow), which rapidly disappears as the field is tilted off the trigonal axis. This field scale is the only feature of our data to be absent in the torque data in this field window.

\subsection{High field measurements}
Fig. 2 presents the Nernst data taken with a rotating set-up at T= 1.5 K in a DC resistive magnet up to 28 T.  As the field is tilted off the trigonal axis, the structure of the Nernst response above 9 T evolves. Panel b of the same figure presents a color map of the Nernst response in the (B, $\theta_{1}$) plane, which clearly exposes the strongly anisotropic field scales discovered by Li \emph{et al.} in their angular-dependent torque magnetometry experiments\cite{li}. These field scales trace quasi-vertical lines in the  (B, $\theta_{1}$) plane. In addition to these two lines, there are three additional field scales above 9 T, which cross the $\theta_{1}=0$ axis close to the position of the three anomalies detected in the earlier Nernst experiment with a field nominally along the trigonal axis, but with no \emph{in situ} control of orientation\cite{behnia1}. The detection of these two distinct field scales by the same experiment definitely rules out the misalignment scenario proposed\cite{sharlai} as an explanation for the unexpected Nernst anomalies\cite{behnia1}. In this spectrum, the two quasi-vertical lines correspond to the passage of an electron ellipsoid sub-level (labeled 0$^{+}_{e}$) through the chemical potential as the field is tilted\cite{alicea,sharlai}. First detected by torque magnetometry\cite{li}, they are also visible in angular-dependent magnetoresistance data\cite{fauque1}.

We also used a two-axis rotation set-up, which allowed us to delimit the triangular around the trigonal axis formed by the intersection of each of the three 0$^{+}_{e}$ sub-levels with the Fermi level. As seen in Fig. 3, even at 28 T, this triangle has a finite size allowing us to identify the trigonal axis. We checked the presence of the three ultraquantum Nernst peaks in such a configuration with a virtually perfect alignment of the magnetic field. Their angular dependence of these peaks is much weaker that the two quasi-vertical lines which delimit the central region. This may suggest an explanation for their absence in the torque data\cite{li}. There is no torque when the field is strictly aligned along the trigonal axis. When the field is tilted, the torque response, which is proportional to the anisotropy of the magnetic susceptibility, is dominated by the more anisotropic field scales. As pointed out by Li and co-workers\cite{li}, in a torque study sweeping $\theta_{1}$, the ellipsoid $e3$ of Fig.1a, for example, is invisible because of its negligible contribution to anisotropic magnetization.

A striking feature of the Fig. 2b and Fig. 3b, is the absence of mirror symmetry between positive and negative $\theta_{1}$. In Fig. 2b, the line crossing  $\theta_1$=0 at 18 T does not respect the reflection symmetry of the crystal. Moreover, in Fig. 3b, the legs of the triangle delimiting the central region around the trigonal axis do not display the same intensity. It is unlikely that these features arise because of a residual imperfection in controlling the field orientation, in particular in the latter case, as the two axis set-up scans a solid angle.

Fig. 4 presents the data obtained at different temperatures when the field is aligned along the trigonal axis. The thermal evolution of the quantum oscillations confirms the identification of Nernst peaks according to their angular dependence. The hole peaks are more robust than the electron peaks as the sample is warmed up. At T=4.3 K, while the electron peaks are already smeared out, hole peaks persist.  The persistence of the hole peaks up to higher temperatures is in agreement with their lighter effective mass in this configuration ($m^{\|}_{h}\simeq 0.064 m_{0}$ for holes and $m^{\|}_{e}\simeq 0.26 m_{0}$\cite{smith}). The lower panel presents the high-field Nernst signal as a function of the inverse of the magnetic field, with the magnetic field aligned along the trigonal axis with our two-axis set-up. The three ultraquantum anomalies fade away with warming almost at the same temperature as the low-field peaks clearly identified as those associated with electron ellipsoids. At T=5 K, the hole anomalies are still present, but the ultraquantum peaks are all wiped out. This observation points to electron ellipsoids as the source of the three high-field anomalies. This conclusion is confirmed by the size of the latter matched to electron and hole peaks below the quantum limit.

\section{Discussion}
Controlling the orientation of the magnetic field with sub-degree accuracy allows us to confirm the complex theoretical one-particle spectrum of bismuth at high magnetic fields\cite{alicea,sharlai} in its basic lines. For the first time, Nernst peaks associated with the very anisotropic electron ellipsoids can be clearly identified thanks to their sharp angular variation, which is in very good agreement with theory\cite{sharlai,alicea} as well as the torque data by Li and co-workers\cite{li}. The second and more important conclusion of this study is that the three high-field Nernst peaks reported previously are not one of these anomalies caused by the intersection of a Landau level and the chemical potential.

It has been recently pointed out that when the field is tilted off the trigonal axis, the large electron-phonon coupling can lead to redistribution  of carriers between pockets and sharp transport features may arise even in the one-particle picture\cite{littlewood}. Therefore, the presence of the anomalous Nernst peaks with a virtually perfect alignment between the magnetic field and the trigonal axis is particularly significant. As seen in Fig. 4, the Nernst peaks resolved in this configuration smoothly vanish with warming and there is no detectable critical temperature. Moreover, their field position does not shift with temperature. These peaks are not signatures of a thermodynamic phase transition such as the one which occurs in graphite, believed to be a field-induced Density-Wave (DW) transition\cite{yaguchi}. Bismuth, in contrast with graphite, keeps its metallicity up to a magnetic field as large as 55 T\cite{fauque2}.

We conclude that at magnetic fields of 12.5 T, 18.2 T and 25.7 T, for which, according to the one-particle picture, the Fermi level is between one occupied and one empty Landau level of all three electron pockets, the Nernst signal peaks as if there was a Landau level crossing. This is also the case of the peak at 6.9 T (green symbols in Fig.1 and Fig.4a). We also note that the large Nernst peak of 38 T, resolved previously and associated with the hole pocket\cite{fauque2}, occurs when according to the theoretical one-particle spectrum, the chemical potential is between the occupied 0$^{-}_{h}$ and the empty 1$^{-}_{h}$ Landau level. What occurs at these fields is  not a thermodynamical phase transition associated to a symmetry breaking order parameter, but a topological\cite{wen} one, which occurs each time a Landau (sub-)level intersects the Fermi level\cite{blanter}.

When the bottom of a Landau level intersects the Fermi level, there is a sharp enhancement in entropy per carrier and thus a Nernst peak\cite{zhu,bergman}. In the presence of Zeeman coupling, the degeneracy associated with spin degrees of freedom is lifted and sub-levels emerge. The challenge for theory is to find a mechanism to produce additional Landau sub-levels and leading to Nernst peaks at particular magnetic fields. Since the three electron valleys are degenerate when the field is along the trigonal axis, the valley degrees of freedom are a natural direction to look. However, as long as the three valleys are strictly identical, a field along the trigonal axis is not expected to lift this degeneracy. Very recently, a new theoretical scenario invoking spontaneous valley polarization as a result of electron interaction has emerged and may prove to be relevant to our results\cite{abanin}. In particular, this scenario would provide a natural explanation for the absence of rotational symmetry in our high-field data.

In summary, the study of the angular-dependent Nernst response fine tunes the challenge addressed to theory. First of all, the complex theoretical one-particle spectrum of bismuth at high magnetic fields\cite{alicea,sharlai} is confirmed in its basic lines. Nernst peaks associated with the very anisotropic electron ellipsoids are clearly detected and are in very good agreement with both calculations\cite{alicea,sharlai} and the torque data\cite{li}. More importantly, \emph{none} of the three high-field Nernst peaks reported previously\cite{behnia1} are caused by the intersection of a Landau level and the Fermi level in the one-particle spectrum. Finally, the temperature dependence of the anomalous Nernst peaks clearly links them to the electron ellipsoids. The origin of these additional Landau sub-levels, unexpected in the non-interacting picture, remains an open question.

\section{Methods}
\subsection{Samples}

In total, five bismuth single crystals with a length between 2 to 5 mm and a thickness between 0.5 to 2.1 mm were studied . Their Residual Resistivity Ratio (RRR: that is the change in resistance from room temperature to 4.2 K) of these samples ranged from 40 to 120. In all five crystals, for a field aligned with sub-degree accuracy along the trigonal axis, the three ultraquantum Nernst peaks were found to occur at the same magnetic field (within a window of 1 T). The anomalies were more pronounced in the samples with the highest RRR).

\subsection{Angular dependent Nernst measurements}
The Nernst coefficient was measured with a standard one-heater-two-thermometer set-up. Both the heater and the thermometers were RuO$_{2}$ chips attached to the sample through silver wires. The set-up was rotated using a piezoelectric linear positioner provided by the Attocube company (www.attocube.de). A one-axis rotation set-up (allowing a rotation window of 20 degrees) was used to measure the angular-dependent Nernst effect in a dilution refrigerator down to 0.18 K and in presence of a superconducting magnet up to 12 T. The same set-up was then used to study the Nernst effect and its angular-dependence in a DC resistive magnet of LNCMI-Grenoble up to 28 T and down to 1.5 K.

A second set-up allowing rotation in two perpendicular planes, albeit in an angular window restricted to 5 to 7 degrees, was used to obtain the data presented in Fig. 3 and Fig. 4.

The angle between the magnetic field and the sample was determined using a Hall sensor attached to the rotating stage. In the case of the two-axis set-up, we used two perpendicular Hall sensors to determine the two angles. The relative accuracy of angular determination was about 0.05 degree.

\textbf{Acknowledgements:} We thank J. Alicea, L. Balents, A. Millis and Y. Sharlai for stimulating discussions. This work is supported by Agence Nationale de Recherche (ANR-08-BLAN-0121-02) as part of DELICE  and by EuroMagNET II under the EU contract number 228043. B.F. is a Newton International Fellow. Z.Z. acknowledges a scholarship granted by China Scholarship Council.

\textbf{Authors contribution:}  H.Y., B.F., L.M., A.B.A and K.B carried out the high-field measurements. H.Y., B.F. and Z.Z. assisted by K.B. performed the low-field measurements. C.U. made the bismuth single crystals used in this study. All the authors participated in the analysis and discussion of the results. K.B. wrote the paper.

\textbf{Conflict of interest statement:} The authors declare no conflicting financial interests.

\textbf{Corresponding author} Correspondence should be sent to K. B. (kamran.behnia@espci.fr)

\begin{figure}
\resizebox{!}{0.75\textwidth}{\includegraphics{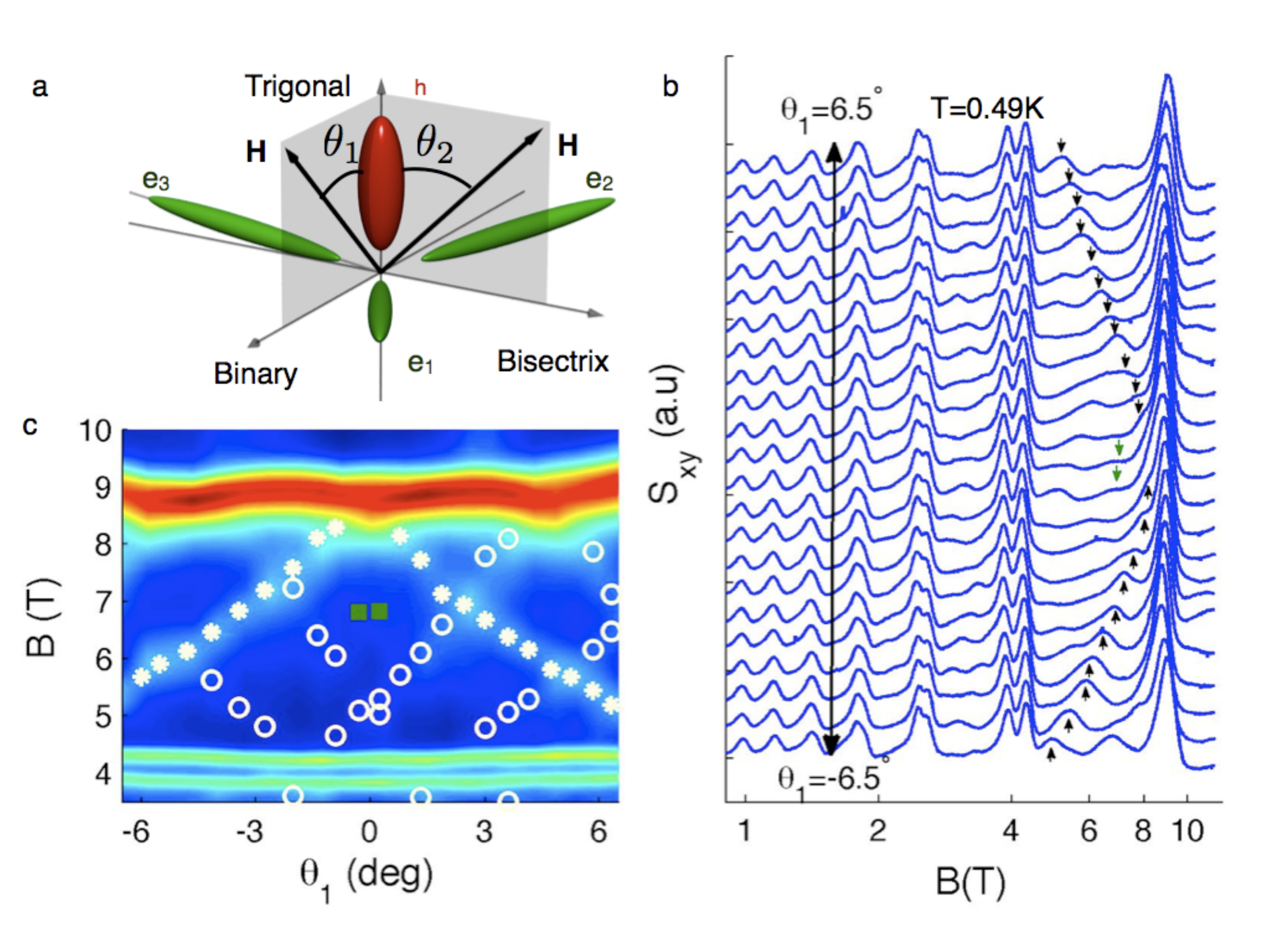}}\caption{ \textbf{Fermi surface, low-field Nernst quantum oscillations and angular dependence of the Landau levels-} a: The Fermi surface of bismuth with the hole pocket in red and electron pockets in green. $\theta_{1}$ ($\theta_{2}$) designate angles between the magnetic field and the trigonal axis in the trigonal-binary (trigonal-bisectrix) rotating plane; b: Quantum oscillations of the Nernst signal up to 12 T with a tilted magnetic field at T=0.49 K. Note the rapid variation of the peak marked by a black arrow;  c: Color plot of the Nernst response in the (B, $\theta_{1}$) plane, where red to blue marks variation from high to low. Superposed symbols mark the position of Nernst maxima associated with three different electron Landau levels.}
\end{figure}
\begin{figure}
\resizebox{!}{1\textwidth}{\includegraphics{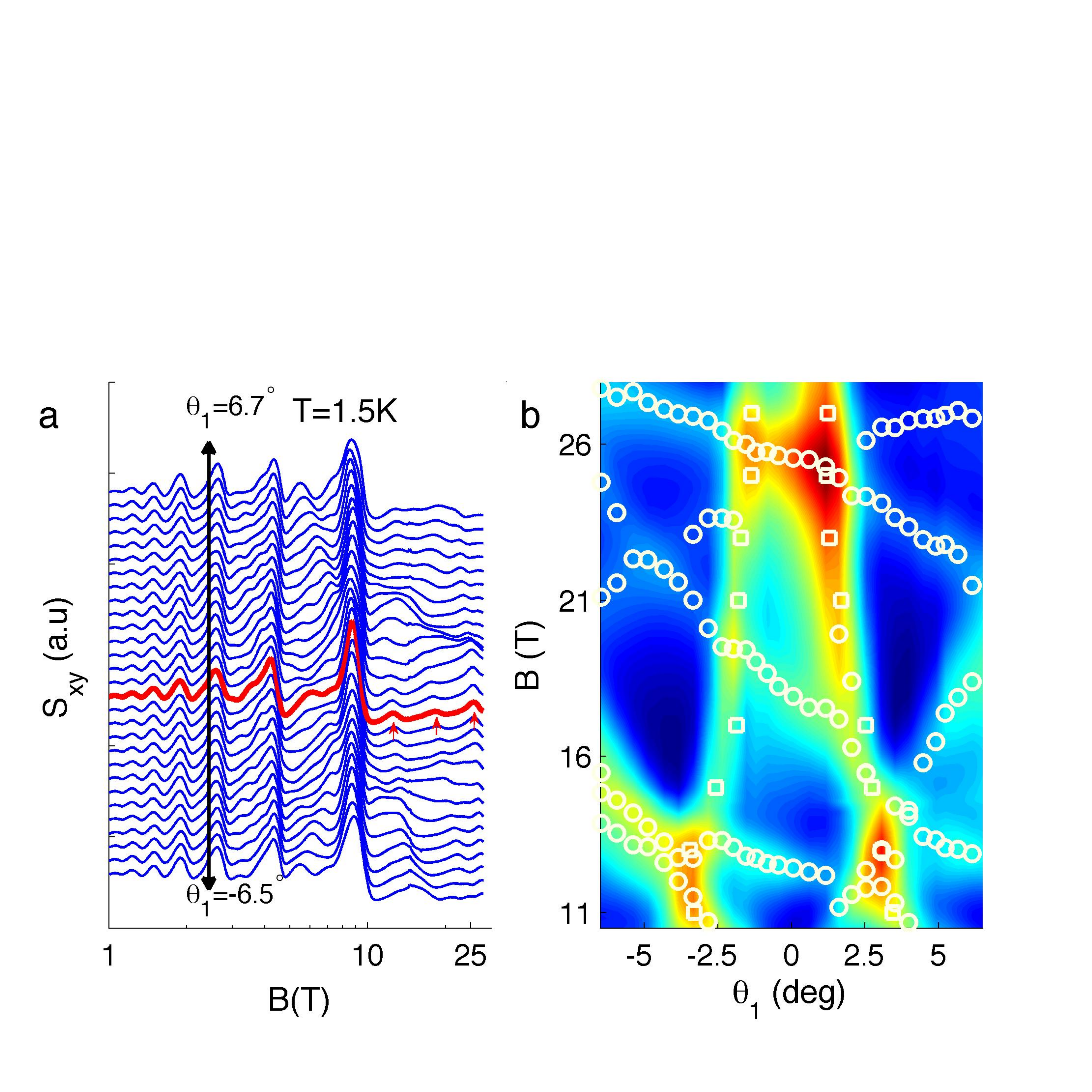}}\caption{\textbf{High-field Nernst quantum oscillations and phase diagram-} a: Field dependence of the Nernst signal up to 28 T for various $\theta_{1}$ tilt angles at T=1.5 K. The  red curve corresponds B $\parallel$ trigonal. The three ultraquantum anomalies are marked with arrows; b: Color plot of the Nernst response in the (B, $\theta_{1}$) plane. Blue to red is minimum to maximum. Nernst peaks resolved in S$_{xy}$ (B) (S$_{xy}(\theta_{1}$)) sweeps are marked by circles (squares).}
\end{figure}
\begin{figure}
\resizebox{!}{0.6\textwidth}{\includegraphics{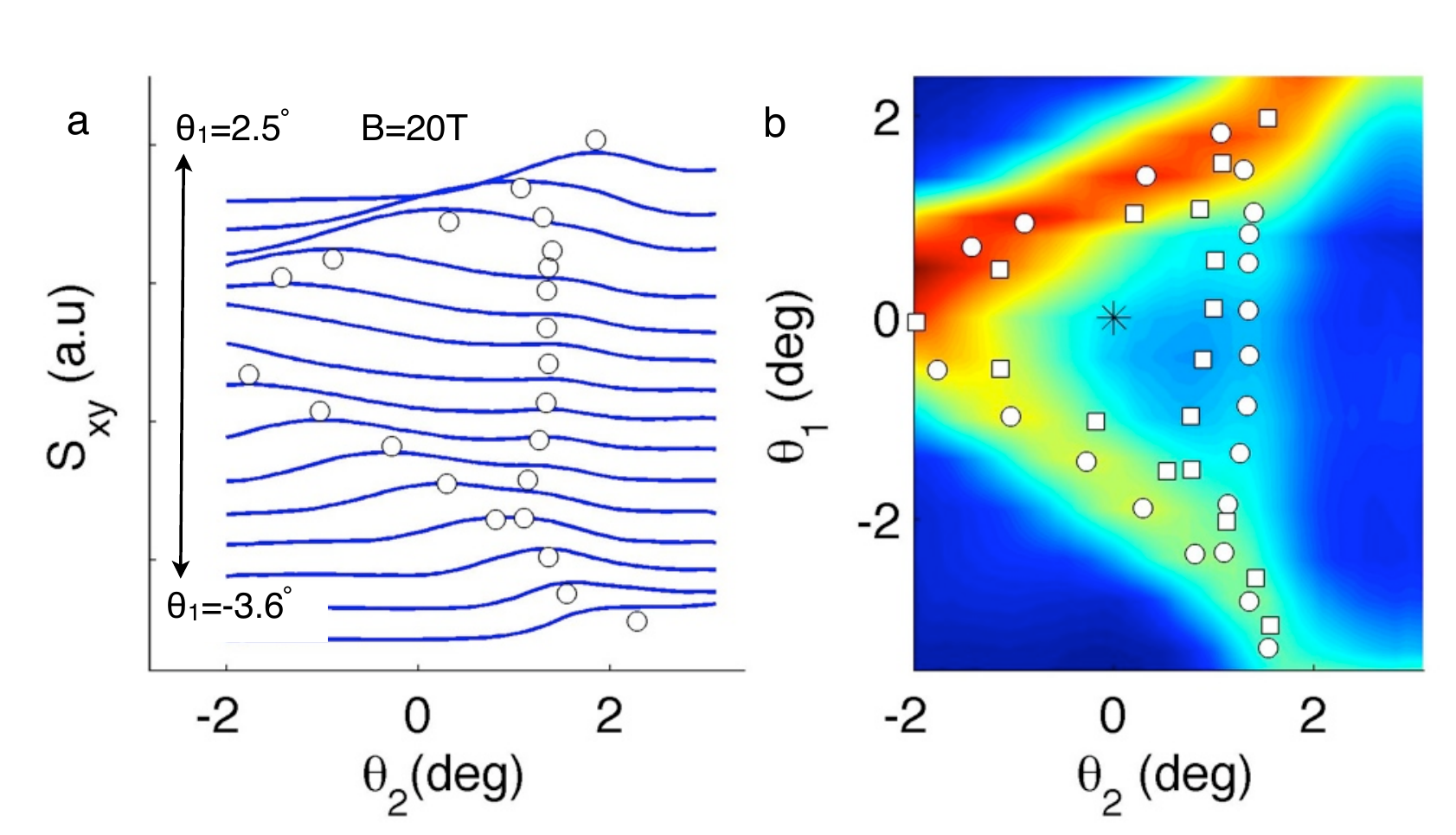}}
\caption{ \textbf{Two-axis rotation set-up data-} a:  $\theta_{2}$ -dependence of the Nernst signal for various  $\theta_{1}$s at B=20 T. Circles follow the shift in the position of the Nernst maxima. b: The angular position of the Nernst maxima in the ( $\theta_{1}$,  $\theta_{2}$ ) plane at 20 T (circles) and 28 T (squares) superposed on a color plot of the Nernst signal at 20 T (blue to red is minimum to maximum).  The trigonal axis is marked by a star.}
\end{figure}
\begin{figure}
\resizebox{!}{1\textwidth}{\includegraphics{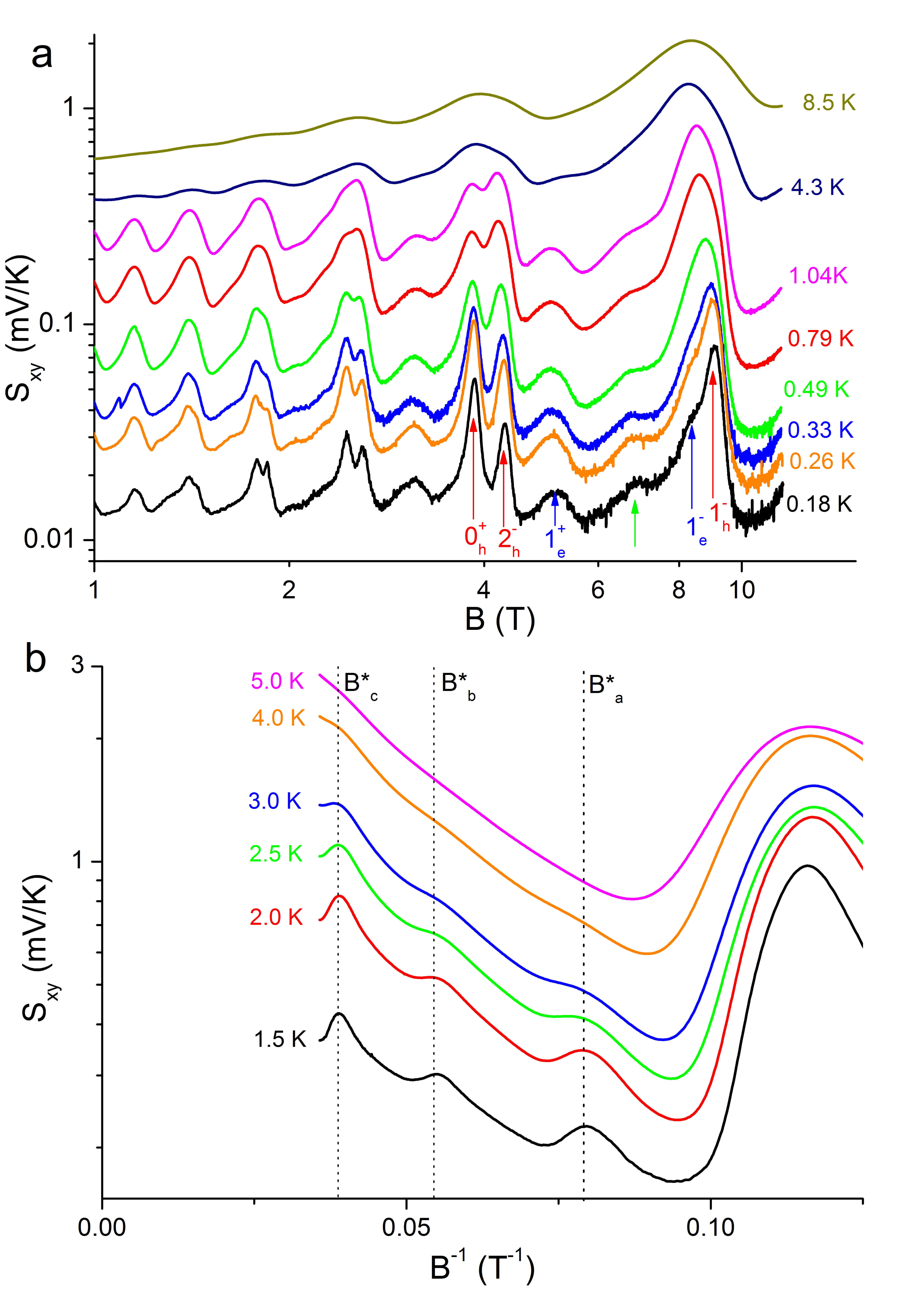}}\caption{\textbf{Thermal evolution of the Nernst quantum oscillations-} a:  Quantum oscillations of the Nernst signal up to 12 T at different temperatures for magnetic aligned along the trigonal axis. Landau levels for electrons and holes are indexed. b: The Evolution of the  high-field Nernst signal as a function of the inverse of the magnetic field with increasing temperature when the magnetic field is aligned along the trigonal axis. The high-field anomalies and the Nernst peaks associated with the electron pockets vanish almost at the same temperature, while the peaks associated with the hole pocket survive.}
\end{figure}
\end{document}